\documentclass[letterpaper,twocolumn,english,aps,pra,showpacs,superscriptaddress]{revtex4-2}
\pdfoutput=1
\usepackage[T1]{fontenc}
\usepackage[latin9]{inputenc}
\usepackage{color}
\usepackage{amsmath}
\usepackage{amssymb}
\usepackage{graphicx}

\makeatletter

\usepackage{ulem}
\usepackage{float}
\usepackage{lipsum}
\usepackage{babel}

\usepackage{bbold}
\usepackage{hyperref}
\usepackage{amsmath}

\begin{document}


\title{Dynamic quantum-enhanced sensing without entanglement in central spin systems}



\author{Wenkui Ding}
\email{wenkuiding@iphy.ac.cn}
\affiliation{Beijing National Laboratory for Condensed Matter Physics, Institute of
Physics, Chinese Academy of Sciences, Beijing 100190, China}
\author{Yanxia Liu}
\affiliation{Beijing National Laboratory for Condensed Matter Physics, Institute of
Physics, Chinese Academy of Sciences, Beijing 100190, China}
\affiliation{Department of Physics, Yunnan University, Kunming 650500, China}
\author{Zhenyu Zheng}
\affiliation{Beijing National Laboratory for Condensed Matter Physics, Institute of
Physics, Chinese Academy of Sciences, Beijing 100190, China}
\author{Shu Chen}
\email{schen@iphy.ac.cn}
\affiliation{Beijing National Laboratory for Condensed Matter Physics, Institute of
Physics, Chinese Academy of Sciences, Beijing 100190, China}
\affiliation{School of Physical Sciences, University of Chinese Academy of Sciences,
Beijing, 100049, China}
\affiliation{Yangtze River Delta Physics Research Center, Liyang, Jiangsu 213300, China}


\begin{abstract}
We propose a dynamic quantum sensing scheme by using a quantum many-spin system composed of a central spin interacting with many surrounding spins.
Starting from a generalized Ising ring model, we investigate the error propagation formula of the central spin and it indicates that Heisenberg scaling can be reached while the probe state only needs to be a product state. Particularly,  we derive an analytical form of the dynamic quantum Fisher information in a limit case, which explicitly exhibits the Heisenberg scaling. By comparing with numerical results, we demonstrate that the general case can be well approximated by the analytical result when the coupling strength among the surrounding spins is much weaker than the coupling strength between the central and surrounding spins.
This analytic result guides us to find the appropriate probe state and the proper measurement time, to achieve the Heisenberg scaling in realistic situations.
Furthermore, we investigate various effects which are important in practical quantum systems, including the central spin Zeeman term, the anisotropy of the hyperfine interaction and the inhomogeneity of the hyperfine coupling strength.
Our result indicates that the dynamic quantum-enhanced sensing scheme seems feasible in realistic quantum central spin systems, like semiconductor quantum dots.
\end{abstract}

\date{\today}
\maketitle

\section{Introduction}

Quantum sensing~\cite{degen2017quantum,budker2007optical} and quantum metrology~\cite{giovannetti2006quantum,giovannetti2011advances,pezze2018quantum} are becoming the frontiers of quantum technologies nowadays.
Specifically, theoretical research on quantum parameter estimation utilizing quantum properties has attracted great attentions recent years.
On one hand, most of the research on quantum metrology has been the employment of entangled probe state to achieve the sub-shot-noise limit~\cite{wineland1992spin,braunstein1992quantum,holland1993interferometric,giovannetti2004quantum,nagata2007beating}.
On the other hand, since the entangled probe state is extremely difficult to generate and very prone to decoherence~\cite{huelga1997improvement,leibfried2005creation,kolody2010phase,escher2011general}, many techniques to realize the quantum-enhanced metrology without entanglement have also been proposed~\cite{entanglement2010tilma,benatti2013sub,braun2018quantum}.
Particularly, the quantum effects of quantum many-body systems are employed to realize metrology schemes without entanglement.
For example, the quantum phase transition has been proposed to realize quantum-enhanced parameter estimation~\cite{zanardi2008quantum,rams2018limits,chu2021dynamic,mishra2021driving}.
Besides, the non-linear many-body Hamiltonian has also been proposed to enhance the measurement precision, even reaching beyond the conventional Heisenberg limit~\cite{luis2007quantum,boixo2007generalized,boixo2008quantum,roy2008exponentially,choi2008bose,napolitano2011interaction}.

The central spin model, which consists of a central spin interacting with many surrounding spins, can be used to describe realistic quantum many-spin systems, such as the semiconductor quantum dots~\cite{merkulov2002electron,hanson2007spins}, the Nitrogen-Vacancy center in the diamond~\cite{childress2006coherent,hanson2008coherent}, etc.
Recently, these solid-state central spin systems have been widely investigated in quantum information and quantum computation~\cite{taylor2003long,hanson2007spins}, as well as quantum metrology and quantum sensing~\cite{goldstein2011environment,cappellaro2012environment,ding2020quantum}.
Particularly, in Ref.~\cite{goldstein2011environment,cappellaro2012environment}, the central spin model has been proposed to achieve the quantum-enhanced sensing.
However, this sensing scheme is still based on the generation of the entangled probe state.
The non-linear interaction between the central spin and the surrounding spins implies that it is possible to realize the quantum-enhanced sensing of the magnetic field without entanglement.
Recently  a dynamic framework for criticality-enhanced quantum sensing in the quantum Rabi model has been proposed \cite{chu2021dynamic}.
This work inspires us to look for a dynamic scheme for the realization quantum-enhanced sensing without resorting to the entanglement. Particularly, we shall explore a dynamic routine to sense the magnetic field in a central spin model.
We discover that the Heisenberg scaling can  be achieved in our sensing scheme, while neither the entangled probe state nor the quantum criticality is required.
A similar model known as ZZXX-model was studied in Ref.~\cite{fraisse2015coherent} via the numerical calculation of the quantum Fisher information, in which
the authors concluded that the Heisenberg scaling cannot be reached when only the central spin was measured  \cite{fraisse2015coherent}.
However, in this work we demonstrate that the Heisenberg scaling can still be reached dynamically by only measuring the central spin, when appropriate probe state and measurement time is applied. Especially, we can obtain an explicit analytic form of the dynamic quantum Fisher information for a limit case of our studied central spin model, which correctly predicts the dynamics of the ZZXX-model when the number of surrounding spins becomes large enough.
Besides, we investigate more realistic central spin systems and our result indicates that  quantum-enhanced magnetometry using our dynamic sensing framework seems feasible.
In particular, our scheme has the great advantage that neither entangled probe state nor quantum criticality is required.

The rest of paper is organized as follows. In Sec.II, we first describe the formalism for the calculation of quantum metrology and introduce our model system. In  Sec.III, we study both the local and global quantum Fisher information for the central spin systems and derive analytical expression of the dynamic quantum Fisher information in a limit case. In Sec.IV, we discuss effects related to more realistic central spin system. A conclusion summary is given in the last section.

\section{Formalism and model}
Before the study of concrete models, firstly we give a brief review of the basics of quantum metrology for the convenience of following calculations.
The quantum fidelity between two quantum states $\hat{\rho}_1$ and $\hat{\rho}_2$ is defined as,
\begin{equation}
\mathcal{F}(\hat{\rho}_1,\hat{\rho}_2)=\text{Tr}(\sqrt{\sqrt{\hat{\rho}_1}\hat{\rho}_2\sqrt{\hat{\rho}_1}}).
\end{equation}
In particular, when the quantum state is continuously dependent on parameter $\lambda$, we can define the fidelity susceptibility~\cite{gu2010fidelity}, which is equivalent to the quantum Fisher information~\cite{braunstein1994statistical},
\begin{equation}
F_\lambda=-4\frac{\partial^2 \mathcal{F}(\rho(\lambda),\rho(\lambda+\delta_\lambda))}{\partial \delta_\lambda^2}|_{\delta_\lambda=0}.
\end{equation}
Specifically, when the quantum state is a pure state, namely $\rho(\lambda)=|\Psi(\lambda)\rangle\langle\Psi(\lambda)|$, we can calculate the quantum Fisher information as follows,
\begin{equation}
F_\lambda=4(\langle \Psi(\lambda)|\overleftarrow{\frac{\partial}{\partial\lambda}}\overrightarrow{\frac{\partial}{\partial\lambda}}|\Psi(\lambda)\rangle-|\langle\Psi(\lambda)|\overrightarrow{\frac{\partial}{\partial \lambda}}|\Psi(\lambda)\rangle|^2).
\end{equation}
Furthermore, for a unitary parameter imprint process~\cite{giovannetti2006quantum,liu2015quantum},
\begin{equation}
|\Psi(\lambda)\rangle=e^{-iH_\lambda t}|\Psi_0\rangle,
\end{equation}
where $H_\lambda=H_0+\lambda H_1$ is a general parameter dependent Hamiltonian, the quantum Fisher information is given by
\begin{equation}
\label{eq:QFI}
F_\lambda=4(\langle \Psi_0|G_\lambda^2|\Psi_0\rangle-|\langle \Psi_0|G_\lambda|\Psi_0\rangle|^2).
\end{equation}
Here, the transformed local generator $$G_\lambda\equiv ie^{iH_\lambda t} \frac{\partial}{\partial\lambda}e^{-iH_\lambda t}$$ can be calculated as follows~\cite{pang2014quantum},
\begin{equation}
\label{eq:TLG}
G_\lambda=\int_0^te^{iH_\lambda t}H_1e^{-iH_\lambda t}ds=-i\sum_{n=0}^\infty\frac{(it)^{n+1}}{(n+1)!}[H_\lambda,H_1]_n,
\end{equation}
where the commutation relation is defined as $[H_\lambda,H_1]_{n+1}=[H_\lambda,[H_\lambda,H_1]_n]$, with $[H_\lambda,H_1]_0=H_1$.

The quantum Fisher information sets a bound to the estimation sensitivity, which is called the Cram\'er-Rao bound~\cite{braunstein1994statistical},
\begin{equation}
\Delta_{\delta_\lambda}(\hat{A},\lambda)\geq F_\lambda^{-\frac{1}{2}}.
\end{equation}
Here, the standard derivation of the measurement value is calculated via the error propagation formula,
\begin{equation}
\label{eq:EPF}
\Delta_{\delta_\lambda}(\hat{A},\lambda)=\frac{\sqrt{\langle \hat{A}^2\rangle_{\rho(\lambda)}-\langle \hat{A}\rangle^2_{\rho(\lambda)}}}{|\frac{\partial\langle \hat{A}\rangle_{\rho(\lambda+\delta\lambda)}}{\partial \delta_\lambda}|_{\delta_\lambda=0}|},
\end{equation}
where $\hat{A}$ is the observable to be measured in the experiment.

We begin with a generalized cental spin model \cite{quan2006decay}
which is described by the Hamiltonian of the Ising ring in a transverse magnetic field interacting with a central spin~\footnote{The spins considered in this article is always spin-$\frac{1}{2}$, so here the Pauli matrix is defined as $\sigma_z=\frac{1}{2}\begin{bmatrix}
1 & 0\\
0 &-1
\end{bmatrix}$, and likewise for $\sigma_x$ and $\sigma_y$.},
\begin{equation}
\label{eq:Ising_model}
H=-J\sum_{i=1}^N\sigma_i^x\sigma_{i+1}^x-h\sum_{i=1}^N\sigma_i^y+A\sigma_0^z\sum_{i=1}^N\sigma_i^z.
\end{equation}
Here, $\sigma_0^{\alpha=x,y,z}$ are the operators of the central spin, $\sigma_{i=1,...,N}^{\alpha=x,y,z}$ are the operators of spins in the Ising ring,
$J$ is the strength of the Ising coupling, and $A$ is the coupling strength between the central spin and surrounding spins in the Ising ring.
Different from the usual central spin model studied in Ref.\cite{quan2006decay}, here we apply the transverse magnetic field along the y-axis, instead of the z-axis coupled with the central spin.
In this work, the strength of transverse magnetic field $h$ is the parameter to be estimated.

\section{Local and global quantum Fisher information}

\subsection{Local quantum Fisher information}
We shall first consider the case with only the central spin being measured.
In order to evaluate the performance of utilizing this parameter-dependent Hamiltonian as the parameterization generator, we will first calculate the error propagation formula when the expectation value of central spin is measured.
Specifically, the initial state (or the probe state) is chosen to be a product state
\begin{equation}
\label{eq:probe_epf}
|\Psi_0\rangle=\frac{1}{\sqrt{2}}(\left|\uparrow\right\rangle+\left|\downarrow\right\rangle)\otimes|\Phi_n\rangle,
\end{equation}
where $\left|\uparrow(\downarrow)\right\rangle$ is the central spin state and $|\Phi_n\rangle$ is the collective state of the spins in the Ising ring.
The time evolution of the system is governed by $|\Psi(t)\rangle=e^{-iHt}|\Psi_0\rangle$ and the central spin expectation value can be expressed as
\begin{equation}
\label{eq:sxt}
\langle \sigma_0^x(t)\rangle=\frac{1}{2}\text{Re}(\langle \Phi_n|e^{iH_+t}e^{-iH_-t}|\Phi_n\rangle),
\end{equation}
where
\begin{equation}
H_{\pm}=-J\sum_{i=1}^N\sigma_i^x\sigma_{i+1}^x-h\sum_{i=1}^N\sigma_i^y\pm \frac{A}{2}\sum_{i=1}^N\sigma_i^z.
\end{equation}
Furthermore, we have the relation
\begin{equation}
e^{-iH_{\pm}t}|\Phi_n\rangle=e^{-i\theta\sum_{i=1}^N\sigma_i^x}e^{-iH_{\text{eff}}^{(\pm)}t}e^{i\theta\sum_{i=1}^N\sigma_i^x}|\Phi_n\rangle,
\end{equation}
where $\theta=\arctan(2h/A)$ and
\begin{equation}
H_{\text{eff}}^{(\pm)}=-J\sum_{i=1}^N\sigma_i^x\sigma_{i+1}^x\pm\sqrt{h^2+\frac{A^2}{4}}\sum_{i=1}^N\sigma_i^z.
\end{equation}

We now consider a specific product probe state given by $|\Phi_n\rangle=|+,+,...+\rangle\equiv|N/2,M_x=N/2\rangle$ with $|+\rangle=1/\sqrt{2}(\left|\uparrow\right\rangle_n+\left|\downarrow\right\rangle_n)$, namely all spins in the Ising ring are polarized along the x-axis.
Now, we have
\begin{equation}
\begin{aligned}
\langle& \Phi_n|e^{iH_+t}e^{-iH_-t}|\Phi_n\rangle= e^{-i\theta N} \times\\
&\langle \frac{N}{2},M_x=\frac{N}{2}|e^{iH_{\text{eff}}^{(+)}t}e^{2i\theta\sum_{i=1}^N\sigma_i^x}e^{-iH_{\text{eff}}^{(-)}t}|\frac{N}{2},M_x=\frac{N}{2}\rangle.
\end{aligned}
\end{equation}

Next, we discuss two important limiting situations.
If $J\gg \sqrt{h^2+{A^2}/{4}}$, then $H_{\text{eff}}^{(+)}=H_{\text{eff}}^{(-)}\approx -J\sum_{i=1}^N\sigma_i^x\sigma_{i+1}^x$, and approximately, $|\Phi_n\rangle=|N/2,M_x=N/2\rangle$ is the eigenstate of $H_{\text{eff}}^{(+)}$ and $H_{\text{eff}}^{(-)}$ with eigenenergy $\epsilon_0$.
This leads to $\langle \Phi_n|e^{iH_+t}e^{-iH_-t}|\Phi_n\rangle\approx 1$, which indicates that no information on the parameter can be retrieved by monitoring the central spin expectation value.
Meanwhile, for the opposite limiting situation with $J\ll \sqrt{h^2+{A^2}/{4}}$, approximately, $H_{\text{eff}}^{(+)}\approx\sqrt{h^2+{A^2}/{4}}\sum_{i=1}^N\sigma_i^z$ and $H_{\text{eff}}^{(-)}\approx-\sqrt{h^2+{A^2}/{4}}\sum_{i=1}^N\sigma_i^z$.
If we choose the evolution time $t=t_0=\pi/\sqrt{h^2+{A^2}/{4}}$, then we have
\begin{equation}
\begin{aligned}
\langle& \Phi_n|e^{iH_+t}e^{-iH_-t}|\Phi_n\rangle\\
&=e^{-i\theta N}\langle \Phi_n|e^{i\pi\sum_{i=1}^N\sigma_i^z}e^{2i\theta\sum_{i=1}^N\sigma_i^x}e^{i\pi\sum_{i=1}^N\sigma_i^z}|\Phi_n\rangle\\
&=e^{-i\theta N}\langle \frac{N}{2},M_x=-\frac{N}{2}|e^{2i\theta\sum_{i=1}^N\sigma_i^x}|\frac{N}{2},M_x=-\frac{N}{2}\rangle\\
&=e^{-2i\theta N}.
\end{aligned}
\end{equation}
Using Eq.(\ref{eq:sxt}), the expectation value of the central spin is given by
\begin{equation}
\begin{aligned}
\langle \sigma_0^x(t_0)\rangle&=\frac{1}{2}\cos(2\theta N)\\
&=\frac{1}{2}\cos[2\arctan(\frac{2h}{A}) N]. \label{sigma0x}
\end{aligned}
\end{equation}
By substituting this result into the error propagation formula in Eq.(\ref{eq:EPF}), we obtain
\begin{equation}
\begin{aligned}
\mathcal{E}_h^{-1}=&\frac{\langle [\sigma_0^x(t_0)]^2\rangle-\langle \sigma_0^x(t_0)\rangle^2}{|\frac{\partial \langle \sigma_0^x(t_0)\rangle}{\partial h}|^2}\\
=&\frac{(A^2+4h^2)^2}{16A^2N^2},
\end{aligned}
\end{equation}
which indicates the Heisenberg scaling with respect to $N$, i.e. the number of spins in the Ising ring.

Since here the measurement is only done on the central spin, now in order to estimate the precision bound of this local measurement, we need to calculate the \textit{local} quantum Fisher information corresponding to the reduced density matrix of the central spin.
We can calculate this quantity by using the formula for the quantum Fisher information of one qubit~\cite{zhong2013fisher}:
\begin{equation}
\label{eq:local_qfi}
F_h^0=
\left\{
\begin{array}{lr}
|\partial_h\mathbf{V}|^2+\frac{(\mathbf{V}\cdot\partial_h\mathbf{V})^2}{1-|\mathbf{V}|^2}, & \text{  if }|\mathbf{V}|<1,\\
|\partial_h\mathbf{V}|^2, & \text{  if }|\mathbf{V}|=1,
\end{array}
\right.
\end{equation}
where $\mathbf{V}=(2\langle \sigma_0^x(t)\rangle,2\langle \sigma_0^y(t)\rangle,2\langle \sigma_0^z(t)\rangle)$ is the spin vector on the Bloch sphere.
Following the same procedure for deriving Eq.(\ref{sigma0x}), we have
\begin{equation}
\langle \sigma_0^y(t_0)\rangle=\frac{1}{2}\sin[2\arctan(\frac{2h}{A}) N],
\end{equation}
and $\langle \sigma_0^z(t_0)\rangle=0$.
Then, by substituting these expectation values and their derivatives into Eq.~\ref{eq:local_qfi}, we get
\begin{equation}
\label{eq:local_qfi_J0}
F_h^0=\mathcal{E}_h=\frac{16A^2N^2}{(A^2+4h^2)^2},
\end{equation}
which happens to be the same form as the reciprocal of the error propagation formula in Eq.~\ref{eq:EPF}.
This indicates that $\sigma_0^x$ is indeed the optimal observable to saturate the sensitivity bound for this specific probe state and local measurement.

In fact, employing the standard Ising Hamiltonian with transverse field (with $A=0$ in Eq.~\ref{eq:Ising_model}) to estimate parameters has been investigated in Ref.~\cite{skotiniotis2015quantum}.
Particularly, the result in that paper demonstrated that quantum-enhanced metrology cannot be realized if the probe state is restricted to be the product state.
Therefore, the interaction of the central spin with the Ising ring introduced in our model is crucial to realize the quantum-enhanced sensing without entanglement.

\begin{figure}
\includegraphics[width=0.5\textwidth]{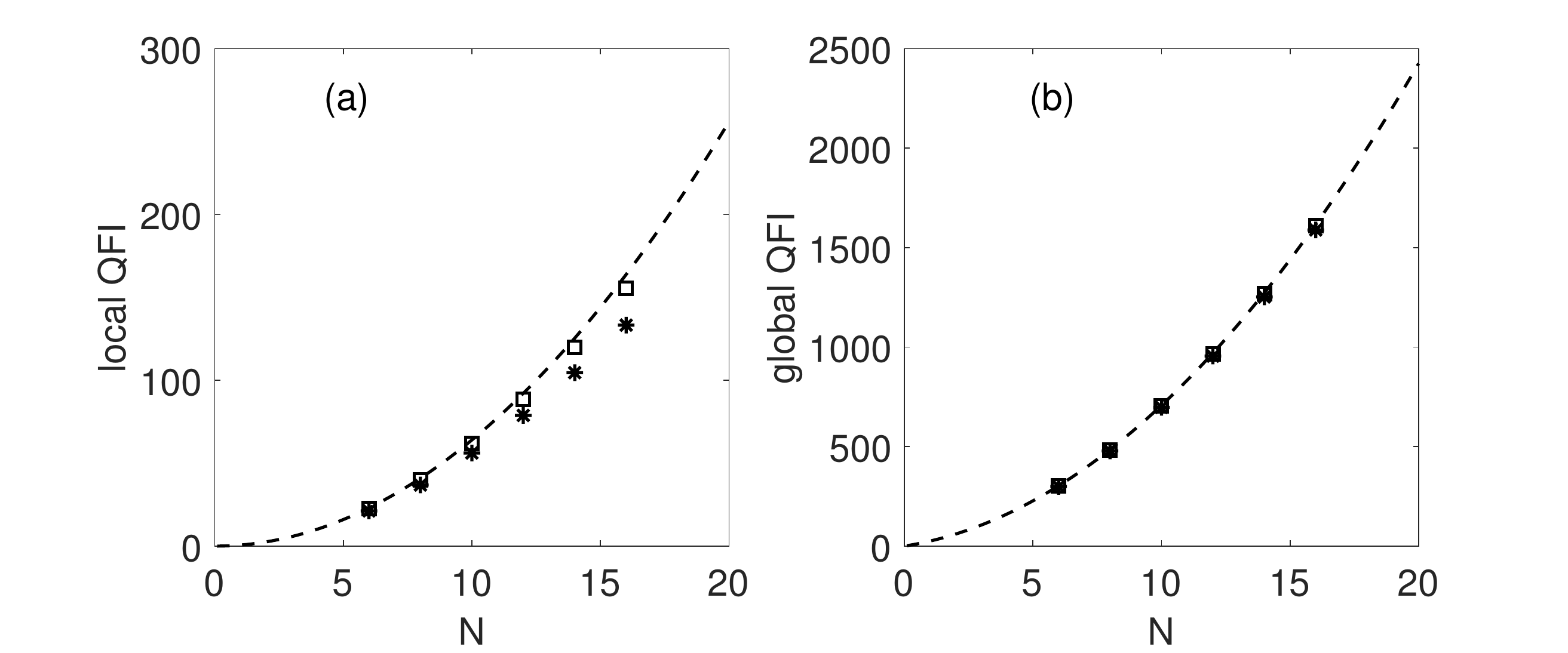}
\centering
\caption{\label{fig:ising_coupling} (a) The scaling of the local quantum Fisher information with respect to $N$ for different Ising coupling strength $J$. The dashed line corresponds to the analytic result for $J=0$ in Eq.~\ref{eq:local_qfi_J0}, while the squares correspond to $J=0.1$ and the stars correspond to $J=0.2$, respectively. (b) The scaling of the global quantum Fisher information with respect to $N$ for different Ising coupling strength $J$. The dashed line corresponds to the analytic result for $J=0$ in Eq.~\ref{eq:global_qfi_J0}, while the squares correspond to $J=0.1$ and the stars correspond to $J=0.2$, respectively. The other parameters in both figures are set to be $A=1$, $h=1$.}
\end{figure}

In order to investigate the effect of finite Ising coupling strength, we numerically calculate the local quantum Fisher information as function of $N$ for different values of $J$.
The result is plotted in Fig.~\ref{fig:ising_coupling}(a) and it indicates that the analytical result for the case of $J=0$  can give good approximation  for the general case when the coupling strength among the surrounding spins ($J$) is much weaker than the coupling strength between the central and surrounding spins ($A$).  As the Ising coupling strength $J$ increases, the Heisenberg scaling deteriorates.
This implies that it is better to measure a magnetic field with large amplitude (which can be realized by adding a large reference magnetic field) to satisfy the condition, $J\ll \sqrt{h^2+{A^2}/{4}}$, for the optimal sensitivity.

\subsection{Global quantum Fisher information\label{sec:dynamic_QFI}}
As discussed above, the calculation of local  quantum Fisher information reveals that
the case of $J=0$ can  realize the quantum-enhanced
sensing without entanglement.
Now we will focus on the case of $J=0$ and use Eq.~\ref{eq:QFI} and Eq.~\ref{eq:TLG} to deduce the analytic form of the \textit{global} quantum Fisher information.
In comparison with the local quantum Fisher information, the global one gives the ultimate sensitivity bound for all possible measurements, namely, not restricted to measurements only on the central spin.

Before we calculate the dynamic quantum Fisher information for the specific case with $J=0$, we have to calculate the transformed local generator using Eq.~\ref{eq:TLG}.
The Hamiltonian of the system now becomes,
\begin{equation}
\label{eq:main}
\begin{aligned}
H&=-h\sum_{i=1}^N\sigma_i^y+A\sigma_0^z\sum_{i=1}^N\sigma_i^z\\
&\equiv -hI_y+AS_zI_z.
\end{aligned}
\end{equation}
Here, since the Hamiltonian does not contain the Ising terms any longer, we have used the collective spin operator $I_{\alpha=x,y,z}\equiv\sum_{i=1}^N\sigma_i^{\alpha=x,y,z}$, and redefined the central spin operator $S_z\equiv\sigma_0^z$.
For convenience, we may call the central spin the \textit{electron spin} and the surrounding spin the \textit{nuclear spin} in the following sections.


We now use Eq.~\ref{eq:TLG} to calculate the transformed local generator.
For the even commutation terms, we have
\begin{equation}
[H,H_1]_{2n}=-A(A^2S_z^2+h^2)^{n-1}(hS_zI_z+AS_z^2I_y),
\end{equation}
while for the odd terms, we have the commutation relation as follows,
\begin{equation}
[H,H_1]_{2n+1}=iA(A^2S_z^2+h^2)^nS_zI_y.
\end{equation}
Then, the transformed local generator to estimate $h$ is calculated as follows,
\begin{equation}
\label{eq:local_generator}
\begin{aligned}
G_h=&ie^{iHt}\frac{\partial}{\partial h}e^{-iHt}\\
=&-tI_y-A\frac{\sin(\Omega t)-\Omega t}{\Omega^3}(hS_zI_z+AS_z^2I_y)\\
&+A\frac{\cos(\Omega t)-1}{\Omega^2}S_zI_x,
\end{aligned}
\end{equation}
where since the central spin $S=1/2$, we have the oscillation frequency $\Omega=\sqrt{A^2S_z^2+h^2}=\sqrt{A^2/4+h^2}$.


Using Eq.~\ref{eq:QFI}, it is easy to verify that the maximized quantum Fisher information is~\citep{pang2014quantum}
\begin{equation}
F_{max}=(E_{max}-E_{min})^2,
\end{equation}
where $E_{max}$ and $E_{min}$ are the maximal and minimal eigenvalues of $G_h$, respectively.
Here for convenience, we denote the transformed local generator in Eq.~\ref{eq:local_generator} as
\begin{equation}
G_h\equiv \alpha I_y +\beta S_zI_z+\gamma S_zI_x.
\end{equation}
Since $[G_h,S_z]=0$, we can separately find the eigenvalues in the $S_z=1/2$ subspace and the $S_z=-1/2$ subspace.
The calculated result of the maximized quantum Fisher information is
\begin{equation}
F_{max}=(4\alpha^2+\beta^2+\gamma^2)I^2,
\end{equation}
where $I=N/2$ and it clearly shows the Heisenberg scaling.
Generally, we have the relation
\begin{equation}
\mathcal{E}_h\leq F_h^0\leq F_{max}.
\end{equation}
The first relation is simply due to the Cram\'er-Rao bound.
The second relation results from the fact that the local quantum Fisher information is obtained by tracing out the degree of freedom of the nuclear spins, which unavoidably leading to the loss of information on the parameter.

To achieve this maximized quantum Fisher information, the probe state in Eq.~\ref{eq:QFI} should be~\footnote{due to the degeneracy of the spectrum of $G_h$, the choice of the optimal probe state is actually not unique.},
\begin{equation}
|\Psi_0\rangle=\frac{1}{\sqrt{2}}(|E_{max}\rangle+|E_{min}\rangle).
\end{equation}
The eigenstate corresponding to the maximal eigenvalue is
\begin{equation}
|E_{max}\rangle=|\uparrow\rangle\otimes(e^{-iI_z\phi}e^{-iI_y\theta}|I,M_z=I\rangle),
\end{equation}
and the eigenstate corresponding to the minimal eigenvalue is
\begin{equation}
|E_{min}\rangle=|\uparrow\rangle\otimes(e^{-iI_z\phi}e^{-iI_y\theta}|I,M_z=-I\rangle),
\end{equation}
with $\theta=\arctan(\frac{\sqrt{4\alpha^2+\gamma^2}}{\beta})$ and $\phi=\arctan(\frac{2\alpha}{\gamma})$, where $|I,M_z\rangle$ is the eigenstate of $I_z$.
This probe state $|\Psi_0\rangle$ is generally an entangled state, however, what interests us most is whether the Heisenberg scaling still maintains when the probe state is restricted to be a product state~\cite{boixo2008product}.

Similar to the dynamical sensing scheme in Ref.~\cite{chu2021dynamic}, we emphasize on the discussion of the dynamic quantum Fisher information at a specific sensing time $t=t_0=2\pi/\Omega$.
Now the transformed local generator in Eq.~\ref{eq:local_generator} becomes
\begin{equation}
\begin{aligned}
G_h&=(\frac{\pi}{2}\frac{A^2}{\Omega^3}-\frac{2\pi}{\Omega})I_y+\frac{2\pi Ah}{\Omega^3}S_zI_z\\
&\equiv \alpha_0 I_y+\beta_0 S_z I_z.
\end{aligned}
\end{equation}
Here, we consider a specific product probe state,
\begin{equation}
\label{eq:probe_qfi}
|\Psi_0\rangle=\frac{1}{\sqrt{2}}(\left|\uparrow\right\rangle+\left|\downarrow\right\rangle)\otimes|I,M_z=I\rangle,
\end{equation}
where $|I,M_z=I\rangle=\left|\uparrow_n,\uparrow_n,\cdots,\uparrow_n\right\rangle$.
It is readily to verify $\langle \Psi_0|G_h|\Psi_0\rangle=0$, and using Eq.~\ref{eq:QFI} we obtain
\begin{equation}
\label{eq:global_qfi_J0}
F_h=4(\langle \Psi_0|G_h^2|\Psi_0\rangle)=2\alpha_0^2I+\beta_0^2I^2.
\end{equation}
This again obviously manifests the Heisenberg scaling with respect to the number of nuclear spins ($I=N/2$).

It needs to be mentioned that the probe state (Eq.~\ref{eq:probe_qfi}) used to achieve the Heisenberg scaling in $F_h$ is different from the probe state (Eq.~\ref{eq:probe_epf}) used to calculate the error propagation formula in the previous subsection.
This is due to the fact that the global quantum Fisher information discussed in this subsection corresponds to the ultimate bound that the measurement on the observable can be done globally, not limited to the central spin only.

Similar to the discussions for the local quantum Fisher information, here we also numerically investigate the effect of finite Ising coupling strength ($J\neq 0$) to the global quantum Fisher information.
In Fig.~\ref{fig:ising_coupling}(b), we illustrate the scaling of the global quantum Fisher information with respect to $N$ for different Ising coupling strength $J$.
Compared to the local quantum Fisher information in Fig.~\ref{fig:ising_coupling}(a), the scaling of the global quantum Fisher information is less sensitive to the strength of Ising coupling, and the analytic result for $J=0$ in Eq.~\ref{eq:global_qfi_J0} approximates very well with the numerical result when the coupling strength between the nuclear spins is much weaker than the coupling strength among the surrounding spins.
This is reasonable, since the local quantum Fisher information corresponds to the reduced state of central spin, while the finite Ising coupling will accelerate the decoherence of the central spin, the decoherence will unavoidably lead to loss of information on the parameter.
Meanwhile, since the global quantum Fisher information corresponds to the quantum state of the composite system (central spin and surrounding spins), the Ising coupling between the nuclear spins may not lead to loss of the parameter information encoded in the total quantum state.

\section{Realization of the protocol in realistic quantum systems}
\subsection{Effect of the electronic Zeeman term\label{sec:electron_Zeeman}}
In the above section, the coupling between the central spin and the surrounding spins may be implemented by the interaction of the light with two-level atoms, where the interaction of the central spin with the field is neglected.
However, it becomes necessary to take into account the electronic Zeeman term if we want to implement our dynamic sensing scheme using solid-state spin systems, like semiconductor quantum dots.
By taking into account the electronic Zeeman term, the Hamiltonian of the system now becomes
\begin{equation}
\label{eq:include_zeeman}
\begin{aligned}
H&=-h(\sigma_0^y+\sum_{i=1}^N\sigma_i^y)+A\sigma_0^z\sum_{i=1}^N\sigma_i^z\\
&\equiv -h(S_y+I_y)+AS_zI_z,
\end{aligned}
\end{equation}
which corresponds to the ZZXX-model numerically investigated in Ref.~\cite{fraisse2015coherent}.
In this subsection we will show that, for the dynamic sensing scheme proposed in this paper, the dynamics and the dynamic quantum Fisher information corresponding to this Hamiltonian, can actually be well approximated by the Hamiltonian in Eq.~\ref{eq:main}, where the electronic Zeeman term is neglected.

\begin{figure}
\includegraphics[width=0.5\textwidth]{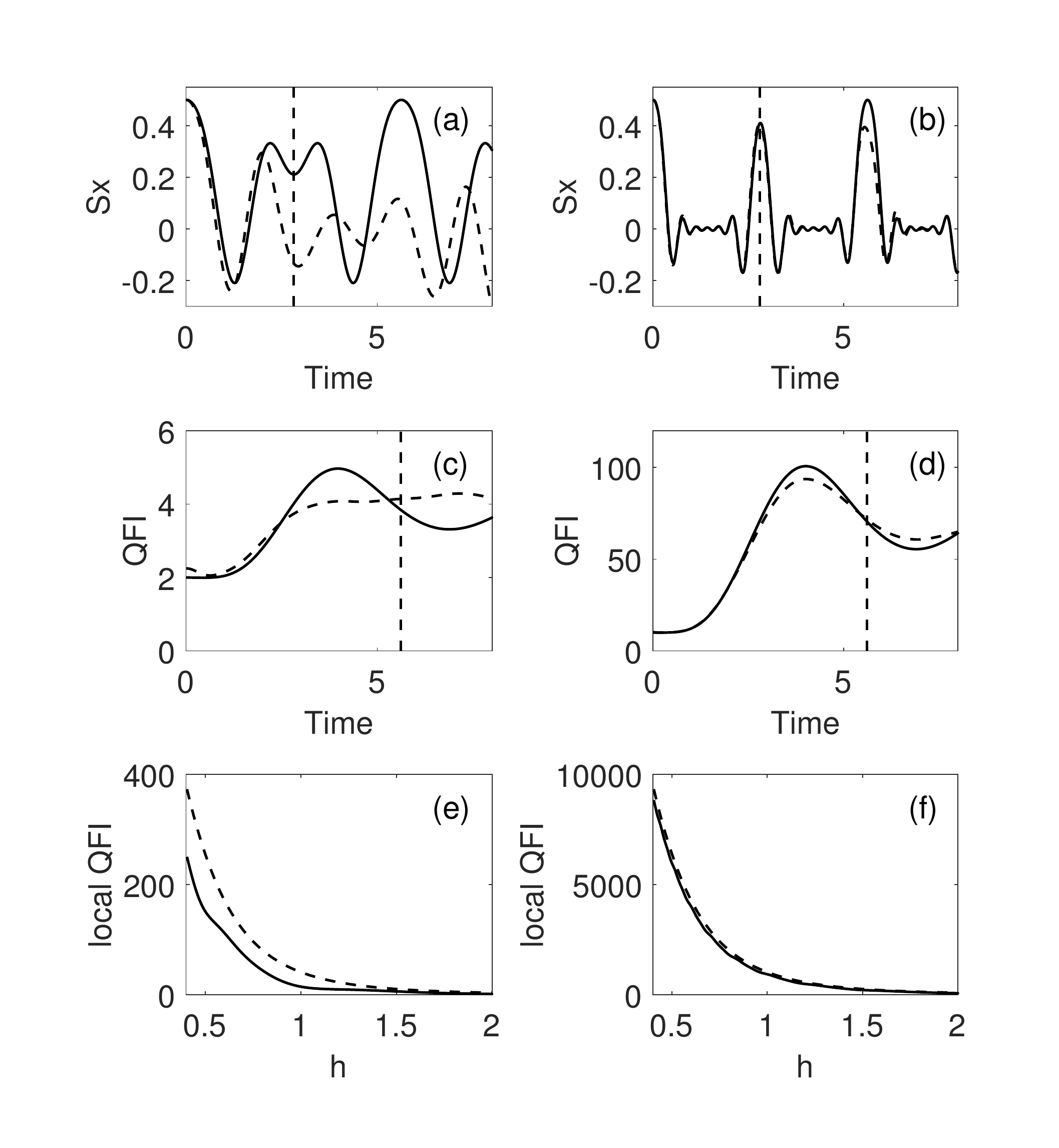}
\centering
\caption{\label{fig:check_approx} The effect of the electronic Zeeman term on the performance of the dynamic sensing scheme. In all figures, the solid lines correspond to the Hamiltonian that taking into account the electronic Zeeman term, while the dashed lines corresponds to the analytic result that neglecting this term. The left column [(a),(c),(e)] corresponds to $N=8$ and the right column [(b),(d),(f)] corresponds to $N=40$. The time evolution of the central spin expectation value $\langle S_x(t)\rangle$ is compared in (a) and (b), with $A=1$ and $h=1$. The time evolution of the global quantum Fisher information is compared in (c) and (d), with $A=1$ and $h=1$, where the vertical dashed line in these figures corresponds to the measurement time $t_0$ deduced from the analytic result.
In (e) and (f), we plot the local quantum Fisher information for the central spin as a function of the external field $h$, while the coupling strength $A=1$ and the measurement time $t=t_0$.}
\end{figure}

As shown in Fig.~\ref{fig:check_approx}, the performance of this approximation (by neglecting the electronic Zeeman term) is actually dependent on $N$, the number of nuclear spins.
This is reasonable, since for our dynamic sensing scheme, the central spin is initialized along the x-axis and then it will precess about the effective magnetic field, from the view of semi-classical picture.
This effective magnetic field consists of the magnetic field $h$ along the y-axis and the nuclear field, $B_{nuc}\propto A\langle I_z\rangle$, along the z-axis.
As the nuclear spin number $N$ increases, the nuclear field $B_{nuc}$ becomes significantly larger than the magnetic field $h$ and the approximated analytic result becomes even closer to the exact numerical result.
Thus, in realistic solid state central spin system, which contains a large number of nuclear spins, the analytic result obtained by neglecting the electronic Zeeman term can be safely utilized in our dynamic sensing scheme.
In other words, when the dynamic sensing scheme by employing the Hamiltonian in Eq.~\ref{eq:include_zeeman} is applied, our analytic result can be used to determine the appropriate probe state, to devise the proper measurement strategy or to estimate the overall sensitivity.

\subsection{Effect of the anisotropy of the hyperfine interaction}
In many realistic quantum central spin systems, the XX-term of the hyperfine interaction may not be neglected, which is described by the so-called XXZ central spin model~\cite{he2019exact}.
The Hamiltonian for such a general central spin system with homogeneous coupling is as follows,
\begin{equation}
\label{eq:hf_anisotropy}
\begin{aligned}
H&=\mathbf{h}\cdot(\mathbf{\sigma}_0+\sum_{k=1}^N\mathbf{\sigma}_k)+\sum_{k=1}^N[\frac{\Delta}{2}(\sigma_0^+\sigma_k^-+\sigma_0^-\sigma_k^+)+A\sigma_0^z\sigma_k^z]\\
&=\mathbf{h}\cdot(\mathbf{S}+\mathbf{I})+\frac{\Delta}{2}(S^+I^-+S^-I^+)+AS_zI_z,
\end{aligned}
\end{equation}
where $\Delta$ is the hyperfine coupling strength of the XX-term and $I^{\pm}=I_x\pm iI_y$, etc.
When $\Delta=A$, namely no anisotropy in the hyperfine coupling, we can check that $[\mathbf{h}\cdot(\mathbf{S}+\mathbf{I}),H]=0$.
It is easy to verify that $G_h=\mathbf{n}\cdot(\mathbf{S}+\mathbf{I})$ with $\mathbf{n}=\mathbf{h}/h$, which indicates that no quantum enhancement can be obtained when the probe state is restricted to be only a product state.
Therefore, the anisotropy in the hyperfine interaction is necessary to achieve the quantum-enhanced sensing without entanglement for our dynamic sensing scheme.
Furthermore, when $\Delta\neq A$ and the magnetic field is applied along the z-axis, it is readily to verify that $G_h=S_z+I_z$ and again no quantum enhancement is possible when the initial state is restricted to be a product state.
Thus, in order to obtain the quantum-enhanced sensing without entanglement, first, there should be anisotropy in the hyperfine coupling; second, the magnetic field to be estimated cannot be applied along the z-axis.

\begin{figure}
\includegraphics[width=0.5\textwidth]{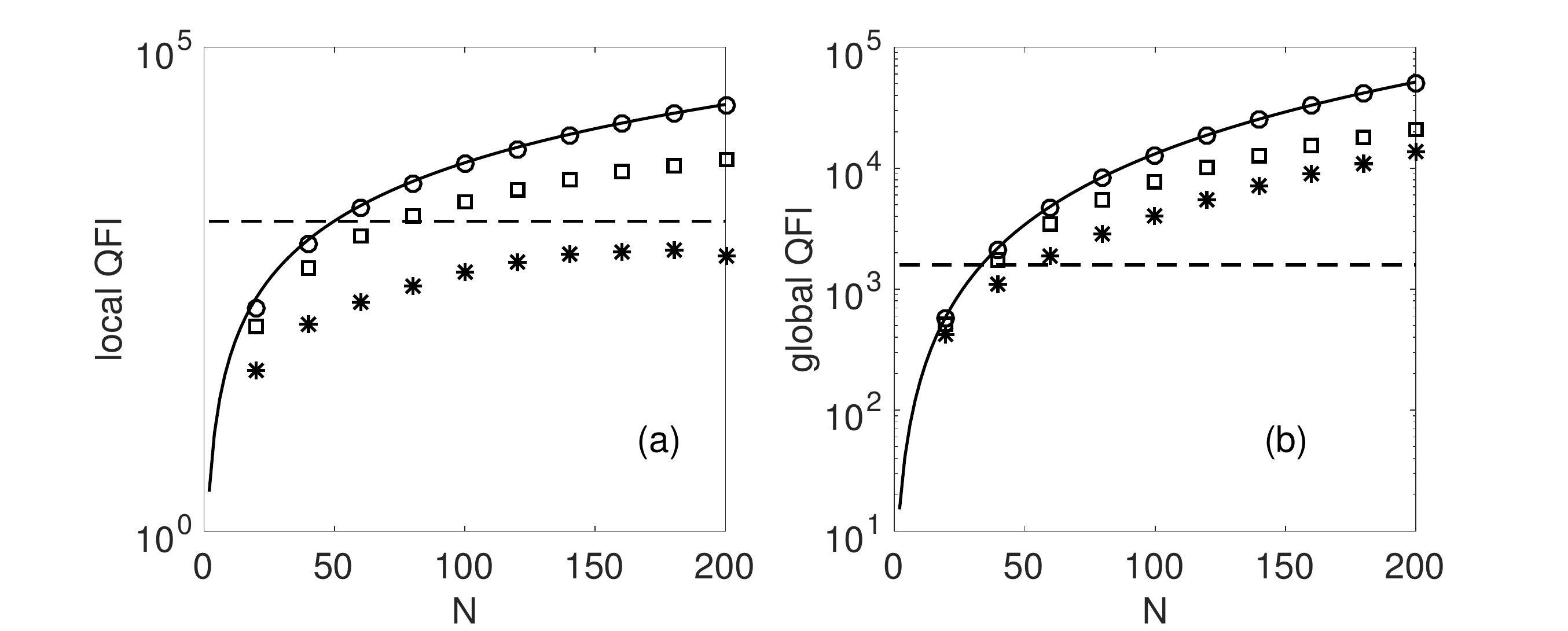}
\centering
\caption{\label{fig:hyperfine_anisotropy} Investigate the effect of the anisotropy of the hyperfine interaction on the performance of the dynamic sensing scheme. The local quantum Fisher information for the central spin is plotted in (a), while the global quantum Fisher information is plotted in (b). In both figures, the solid line corresponds to the analytic result when the electronic Zeeman term is neglected and the hyperfine anisotropy $\Delta=0$. The dashed lines in both figures correspond to the standard quantum limit, which is the ultimate limit when $\Delta=A=0$ in Eq.~\ref{eq:hf_anisotropy} and the probe state is restricted to be the product state. In both figures, the circles correspond to $\Delta=0$, $A=1$ in Eq.~\ref{eq:hf_anisotropy}, while the squares correspond to $\Delta=0.1$, $A=1$ and the stars correspond to $\Delta=0.2$, $A=1$, respectively. }
\end{figure}

We numerically calculate the scaling of the dynamic quantum Fisher information with respect to the nuclear spin number $N$ for various values of $\Delta$ in Fig.~\ref{fig:hyperfine_anisotropy}.
The result indicates that, increasing the anisotropy in the hyperfine coupling will lead to a better performance of our dynamic sensing scheme.

\subsection{Effect of the inhomogeneity of the hyperfine coupling strength}
For realistic solid-state quantum central spin systems, like semiconductor quantum dots, the hyperfine coupling strength between the central electron spin and surrounding nuclear spins is usually inhomogeneous.
The Hamiltonian describing such a system is
\begin{equation}
H=-h(\sigma_{0}^y+\sum_{k=1}^N\sigma_{k}^y)+\sum_{k=1}^NA_k\sigma_{0}^z\sigma_{k}^z,
\end{equation}
where $A_k$ is the coupling strength between the central electron spin and the $k$-th nuclear spin.
Due to the inhomogeneity of the coupling strength ($A_k\neq A$), the collective nuclear spin operator cannot be used any longer and we have to resort to numerics to research the performance of the dynamic sensing scheme.

\begin{figure}
\includegraphics[width=0.5\textwidth]{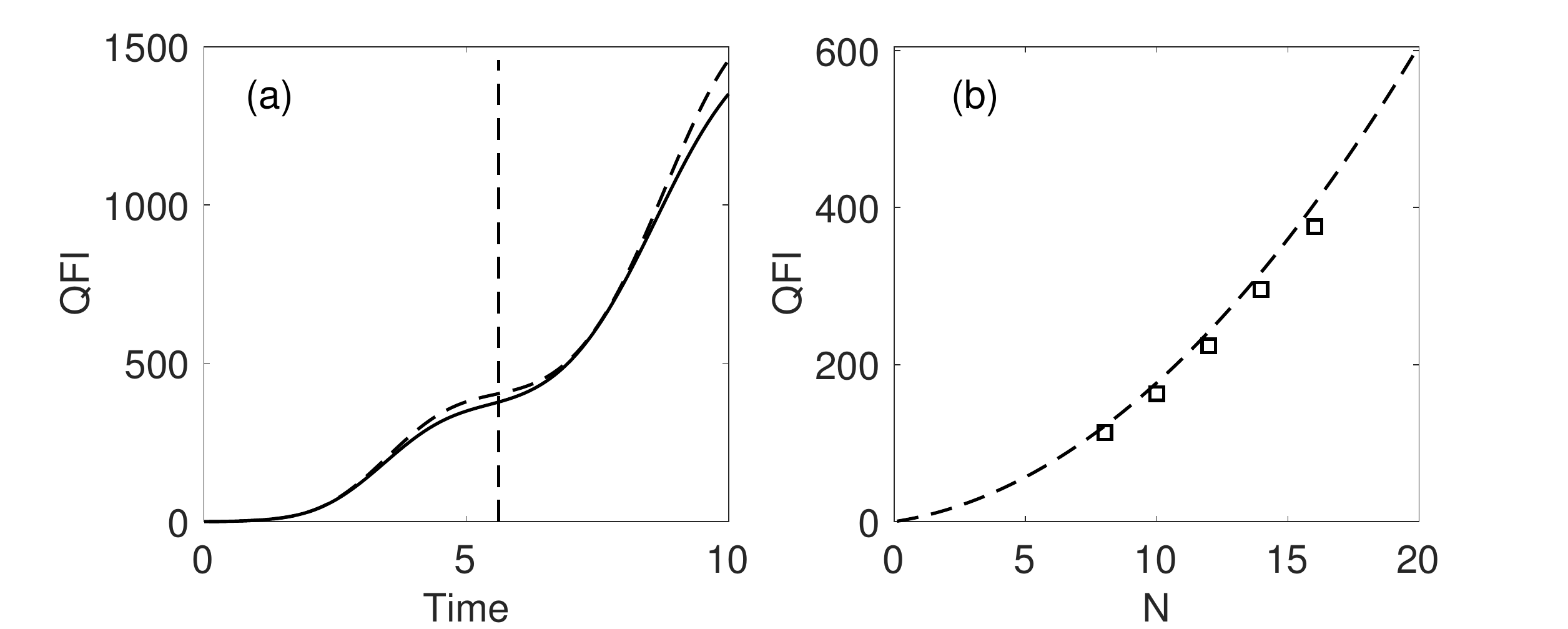}
\centering
\caption{\label{fig:inhomo_qfi} Investigate the performance of the inhomogeneity of the hyperfine coupling strength on the performance of the dynamic quantum sensing. (a). The solid line corresponds to the time evolution of the global quantum Fisher information for $N=16$ nuclear spins with inhomogeneous hyperfine coupling, while the dashed line corresponds to the homogeneous case. The vertical dashed line corresponds to the measurement time. (b). The global quantum Fisher information as a function of the nuclear spin number $N$. The dashed line corresponds to the analytic result of the homogeneous case, while the squares correspond to the numerical result of the inhomogeneous case.}
\end{figure}

Here, we employ the Chebyshev method (by expanding the time evolution operator in terms of Chebyshev polynomials~\cite{dobrovitski2003efficient}) to calculate the dynamics  of this inhomogeneous central spin system and retrieve the scaling of the dynamic quantum Fisher information with respect to $N$.
The result is shown in Fig.~\ref{fig:inhomo_qfi} and it indicates that the performance of such an inhomogeneous sensor can be well approximated by the homogeneous one, as long as the coupling strength in the homogeneous case is set to be $A=\sum_k{A_k}/N$.
Crucially, the Heisenberg scaling still exists for the inhomogeneously coupled central spin system.


%

\subsection{The realistic semiconductor quantum dot system}
We now elaborately consider a practical solid-state quantum central spin system, namely the semiconductor quantum dots, which has been widely researched in the area of quantum computation~\cite{taylor2003long,hanson2007spins} and quantum sensing~\cite{goldstein2011environment,cappellaro2012environment,ding2020quantum} recent years.
The detailed Hamiltonian describing such a quantum many-spin system is as follows~\cite{slichter2013principles,hanson2007spins},
\begin{equation}
H=H_{\text{D}}+H_{\text{Z}}+H_{\text{HF}},
\end{equation}
with,
\begin{equation}
\begin{aligned}
H_{\text{D}}&=\frac{\gamma_n^2}{2}\sum_{j=1}^N\sum_{k=1}^N[\frac{\mathbf{I}_j\cdot\mathbf{I}_k}{r_{jk}^3}-\frac{3(\mathbf{I}_j\cdot\mathbf{r}_{jk})(\mathbf{I}_k\cdot\mathbf{r}_{jk})}{r_{jk}^5}],\\
H_{\text{Z}}&=\mathbf{h}\cdot(\gamma_e\mathbf{S}+\gamma_n\sum_{k=1}^N\mathbf{I}_k),\\
H_{\text{HF}}&=\sum_{k=1}^NA_k[\frac{\delta}{2}(S_+I_{k-}+S_-I_{k+})+S_zI_{kz}],\\
\end{aligned}
\end{equation}
where $\mathbf{S}$ is the electronic spin operator and $\mathbf{I}_k$ is the spin operator of the $k-$th nuclei in the quantum dot.
$H_\text{D}$ describes the dipole-dipole interaction between nuclear spins, where $\mathbf{r}_{jk}$ is the position vector from the $j$-th nuclei to the $k-$th nuclei and $\gamma_n$ is the gyromagnetic ratio of the nuclear spin.
$H_\text{Z}$ corresponds to the electronic Zeeman term and nuclear Zeeman terms, where $\gamma_e$ is the gyromagnetic ratio of the electron spin.
$H_\text{HF}$ describes the hyperfine interaction between the electron spin and nuclear spins, where $\delta$ stands for the anisotropy in the hyperfine interaction.
The hyperfine coupling strength $A_k\propto |\phi(\mathbf{x}_k)|^2$, where $|\phi(\mathbf{x}_k)|^2$ is the electron density at the site $\mathbf{x}_k$ of the $k$-th nuclear spin.

The dipole-dipole interaction between nuclear spins in the quantum dot resembles the role of the Ising coupling in Eq.~\ref{eq:Ising_model}.
Typically, in semiconductor quantum dots, the hyperfine coupling strength between the electron spin and nuclear spins is several orders of magnitude stronger than the strength of the nuclear dipole-dipole coupling~\cite{urbaszek2013nuclear,slichter2013principles}.
Similar to the discussion on the Ising coupling in the previous section, this indicates that the effect from the nuclear dipolar coupling to the overall sensitivity of the dynamic sensing scheme can be negligible.
Another point that needs to be mentioned is the difference in the gyromagnetic ratio between the electron spin and the nuclear spins, namely $\gamma_e\neq\gamma_n$.
However, the discussions in Sec.~\ref{sec:electron_Zeeman} still applies despite the difference in the gyromagnetic ratio.
This is because for typical semiconductor quantum dots, the number of nuclear spins $N\sim 10^4-10^6$ in a single quantum dot, and the effective nuclear filed experienced by the electron spin suppresses the magnetic field.
Thus, in realistic semiconductor quantum dots, the electronic Zeeman term can be safely neglected to analyze the performance of our dynamic sensing scheme.


\section{Conclusion}
In summary, by studying the Ising ring model interacting with a central spin, we show that the Heisenberg scaling in the error propagation formula can be reached by only measuring the central spin dynamics. While the the dynamic quantum Fisher information for the general case can be numerically calculated, we can obtain an analytical form of the dynamic quantum Fisher information to estimate the magnetic field in a limit case, which gives good approximation for the general case as long as $J\ll \sqrt{h^2+{A^2}/{4}}$ is fulfilled.
This analytic result explicitly manifests that the Heisenberg scaling can be reached with an appropriate product probe state and the proper measurement time.
Furthermore, we gradually investigate more realistic central spin models analytically and numerically and our results indicate that the Heisenberg scaling can still be achieved in practical quantum central spin systems, like semiconductor quantum dots.
Our dynamic sensing scheme processes the great advantage that neither entangled probe state nor quantum phase transition is required to obtain the quantum enhancement.
Our theoretical result paves the way to the implementation of quantum-enhanced magnetometry without entanglement in practical quantum many-spin systems.

\begin{acknowledgments}
The work is supported by National Key National Key
Research and Development Program of China (Grant No. 2021YFA1402104), the  NSFC under Grants No.12174436 
and No.T2121001 and the Strategic Priority Research Program of Chinese Academy of Sciences under Grant No. XDB33000000.
\end{acknowledgments}

\bibliographystyle{apsrev4-1}
%

\end{document}